\documentclass[pre,reprint,showpacs,floatfix]{revtex4-1}

\usepackage{amsmath}
\usepackage{graphicx}

\begin{document}

\title{Packaging stiff polymers in small containers: A molecular dynamics study}

\author{D. C. Rapaport}

\email{rapaport@mail.biu.ac.il}

\affiliation{Department of Physics, Bar-Ilan University, Ramat-Gan 52900, Israel}

\date{September 15, 2016}

\begin{abstract}

The question of how stiff polymers are able to pack into small containers is
particularly relevant to the study of DNA packaging in viruses. A reduced version of the
problem based on coarse-grained representations of the main components of the system --
the DNA polymer and the spherical viral capsid -- has been studied by molecular dynamics
simulation. The results, involving longer polymers than in earlier work, show that as
polymers become more rigid there is an increasing tendency to self-organize as spools
that wrap from the inside out, rather than the inverse direction seen previously. In the
final state, a substantial part of the polymer is packed into one or more coaxial
spools, concentrically layered with different orientations, a form of packaging
achievable without twisting the polymer.

\end{abstract}



\maketitle


Consider a relatively stiff polymer chain being drawn into a small container through a
narrow opening. This represents an idealized version of packing double-stranded (ds) DNA
into the capsid of a large virus such as a bacteriophage \cite{kno09,mar10,spe12,ber14}.
Because the packing density is high and the virus must be able to eject the DNA at the
end of its life cycle, ordered packaging of the polymer would appear advantageous. The
basic physics question is, then, how do polymers pack in the absence of any imposed
guiding mechanism? Experimental evidence suggests that DNA packaging involves organized
large loops \cite{ric73,ear77,cer97,lan06}, and much has been learned from theory and
modeling \cite{pur03,ben13,kin01,spa05,ali06,for06,loc06,pet07,mah13}, but the chains
simulated previously may not have been long enough to reveal what actually occurs.

MD (molecular dynamics) simulation is used here to investigate a simple packing model
comprising a self-avoiding chain of linked spheres, subject to bond and bond-angle
interactions, together with a spherical shell in which there is a small portal through
which the chain enters, pulled inside by a suitable force. The goal is to determine
whether ordered packing occurs spontaneously and if so, the dependence on the model
parameters, principally the ratio of the chain dimensions (i.e., the persistence length)
and the shell size. While this reduced problem omits many aspects of its real-world
counterpart, including structural elements and complex short- and long-ranged
interactions, details of which are not readily determined, any systematic behavior
observed in the simulations could contribute to understanding the phenomenon.


Highly simplified models are essential for condensing entire packaging trajectories into
timescales accessible to MD. A coarse-grained polymer model \cite{rap04} is used that
consists of a chain of $N = 8000$ spheres linked by elastic bonds of limited
extensibility. The excluded-volume soft-sphere (SP) interaction is $U(r) = 4 \epsilon
[(\sigma/r)^{12} - (\sigma/r)^6] + \epsilon$ for sphere separation $r < r_c = 2^{1/6}
\sigma$. Reduced MD units will be used subsequently: the sphere mass is unity, and
length, energy, time and temperature are expressed in terms of $\sigma$, $\epsilon$,
$\sigma / \sqrt{\epsilon}$ and $\epsilon / k_\text{B}$. Adjacent chain spheres are
bonded by a reversed pair of SP interactions with origins separated by 2.1; the measured
mean bond length is $l_b = 1.051$ with a 3\% variation. Chain bending is governed by the
interaction $U(\theta) = 0.5 f_a (\cos \theta - 1)^2$, where $\theta$ is the angle
between adjacent bonds, with a minimum in the linear configuration ($\theta = 0$). The
stiffness, $f_a$, is the only parameter varied in the present study; for $500 < f_a <
5000$ the measured persistence length, $L_p = l_b / (1 - \langle \cos \theta \rangle)$,
ranges from 50 to 150. There is no torsional interaction to counter chain twist
(important in real DNA), an omission that will be seen as fully justified by the
results.

The capsid is represented as a fixed spherical shell with radius $R_s = 20$ which, for
the stiffest chains, is much smaller than $L_p$; a shell wall of thickness $\approx r_c$
is produced by a radial SP interaction originating at $R_s$. If $\sigma = 2.5\text{nm}$,
a representative value for coarse-grained models of ds-DNA, then $R_s = 50\text{nm}$, a
typical capsid size (the $L_p$ range is then 125--375nm);
chain segments forming a loop of this radius would have $\theta
\approx 3^{\circ}$. Several (here 12) fixed spheres are embedded in the shell wall, in
the equatorial plane (normal to the portal axis) and inset by $r_c$. Their task is to
roughen the surface to oppose free rotation of the chain already packed into the shell
during insertion (there is none after insertion ends); in the analogous macroscopic
system \cite{sto11} sliding friction prohibits rotation of this kind.

There is a small circular portal in the shell for chain entry. This hole is effectively
a cylinder with unit radius and half-length, inside which a radial force pulls spheres
into the shell, a simple approximation to nature's ATP-powered motor \cite{kno09}; given
its short length it can hold just two spheres. The force strength (here 1.4) is chosen
to ensure slow, reasonably steady (and mainly unidirectional) transport through the
portal and is the same for all runs. Portal geometry restricts the chain entry direction
to below $45^{\circ}$ from the normal; further reducing this angle using a longer
cylinder would alter the outcome by directing the chain radially, a separate problem not
considered here. Similarly simplified models have been used in earlier
work \cite{spa05,for06,pet07}, although design details and computational methods differ.

Computations are carried out on a massively parallel GPU (graphics processing unit),
where efficiency requires large systems, thus each simulation considers 27 independent
chains simultaneously. The chains are confined to a box of size 360 with rigid walls;
while far smaller than $l_b N$, the size is large enough ($> L_p$) that its effect on
the strongly varying (prior to shell entry) chain conformation is minimal. The initial
state of each chain is a closely spaced helix of radius $\approx R_s$ aligned with the
portal, and the initial ($\approx 40$) spheres are redirected so the first few are
inside the shell. Standard MD methods \cite{rap04} are used, with a time step of 0.005
and a (constant-$T$) thermostat that maintains a temperature of $T=0.4$;
even when the stiffest chain
is bent to fit inside the shell the mean bending energy remains well below the kinetic
energy (0.6) despite $L_p$ being several times $R_s$. Simulations are run for up to $2
\times 10^8$ steps, adequate for complete insertion of most chains. Snapshots of the
coordinates are recorded periodically for analysis.


\begin{figure}
\begin{center}
\includegraphics[scale=0.8]{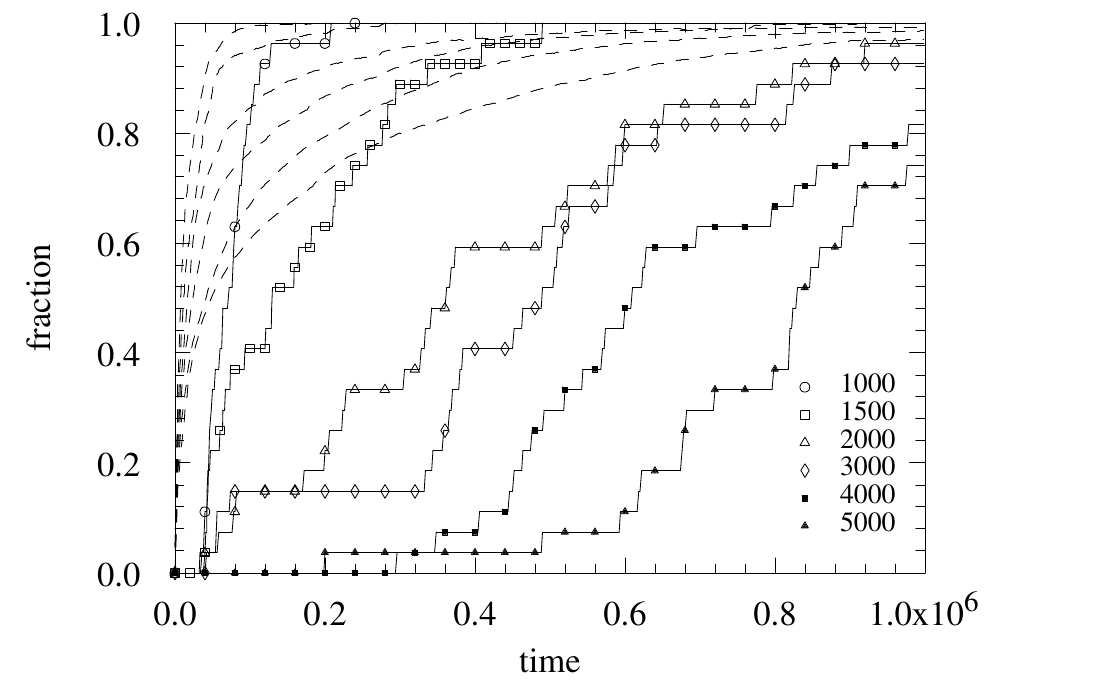}
\end{center}
\caption{\label{fig:1}Fraction of spheres inside shell (dashed lines, increasing $f_a$
from left) and fraction of fully inserted chains vs time (reduced units).}
\end{figure}

Figure~\ref{fig:1} shows how the fractions of spheres inside the shell
(averaged over all chains) and completely
inserted chains depend on time, for several values of $f_a$ covering a range from
moderately flexible to sufficiently stiff chains that entry is seriously impeded. The
mean insertion rate falls as chains become stiffer, a trend that persists to even higher
$f_a$; the typical insertion speed $\approx 0.01$ is just 1\% of $v_\text{therm} =
\sqrt{3 T}$, ensuring near-equilibrium conditions. Individual chain transport (not
shown) can be extremely irregular, as observed experimentally \cite{ber14}. The time
required for complete insertion is also seen to be highly variable.

\begin{figure}
\begin{center}
\includegraphics[scale=0.8]{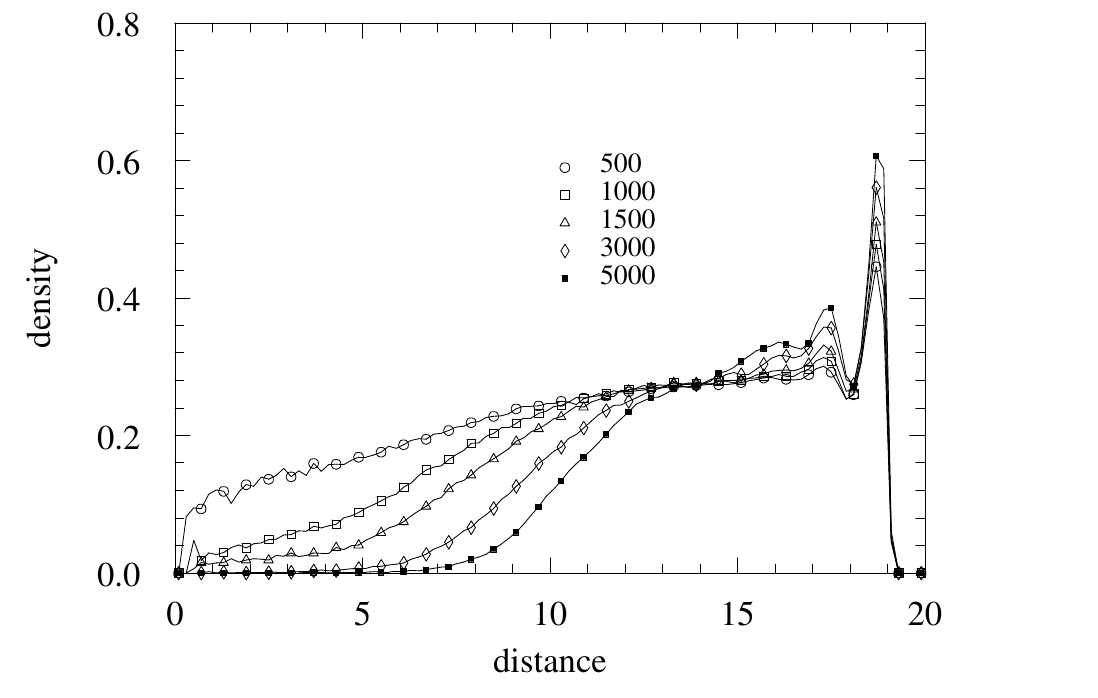}
\end{center}
\caption{\label{fig:2}Radial density distributions (reduced units) for several $f_a$
values.}
\end{figure}

Radial density distributions, each averaged over all fully inserted chains and over 10
snapshots, are shown in Fig.~\ref{fig:2}. Depending on $f_a$, two or three peaks occur
near multiples of $r_c$ from the shell boundary, a signature of layering. Density drops
towards the shell center, with an almost empty sphere of radius $\approx R_s / 2$ for
$f_a = 5000$, and increasingly broad distributions at lower $f_a$. Since the total chain
volume, when approximated by a tube, is $N \pi r_c^2 \approx 6000$, ideally it would
fill $\approx 20\%$ of the shell volume ($4 \pi / 3\, R_s^3 \approx 32000$) when tightly
packed, but Fig.~\ref{fig:2} shows that even the stiffest chains actually expand to
occupy 80\% of the volume because the bending energy involved is small.

\begin{figure}
\begin{center}
\includegraphics[scale=0.8]{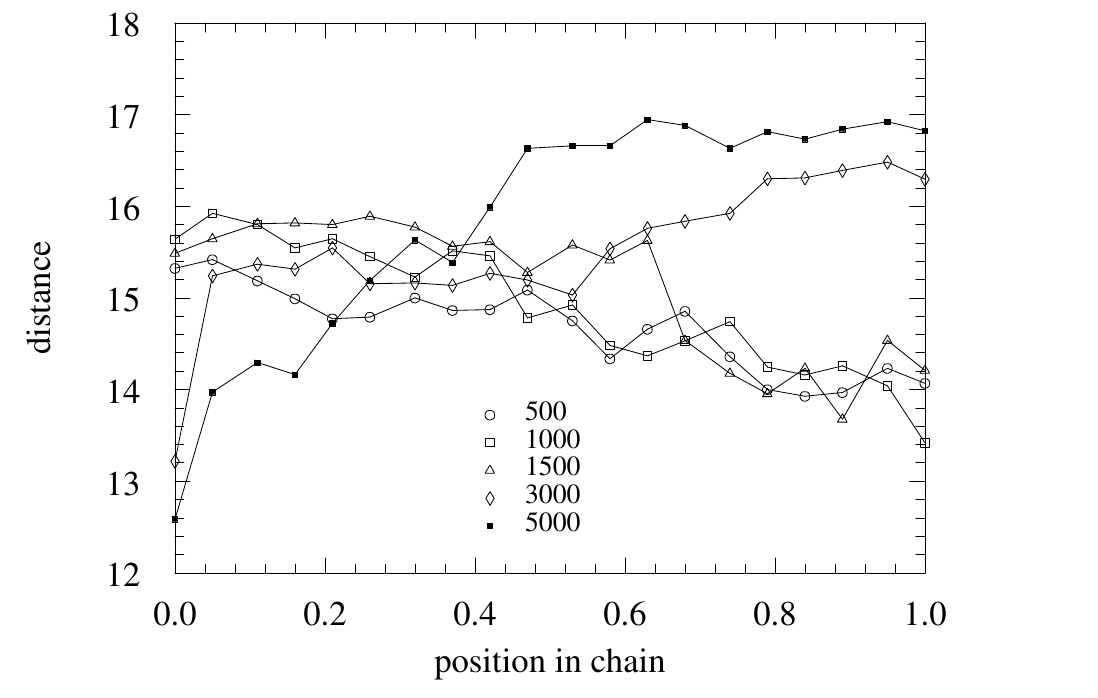}
\end{center}
\caption{\label{fig:3}Mean distance of chain spheres from the shell center as a function
of (normalized) location along chain contour.}
\end{figure}

The question of where chain spheres are positioned is partially answered by
Fig.~\ref{fig:3} which shows their mean distance from the shell center as a function of
location along the chain contour (averaged as before). The preferred packing direction
changes with chain stiffness: for larger $f_a$ chain segments that enter later lie on
the outside (on average), while for smaller $f_a$ the opposite is true, although the
trend is weaker.

\begin{figure}
\begin{center}
\includegraphics[scale=0.23]{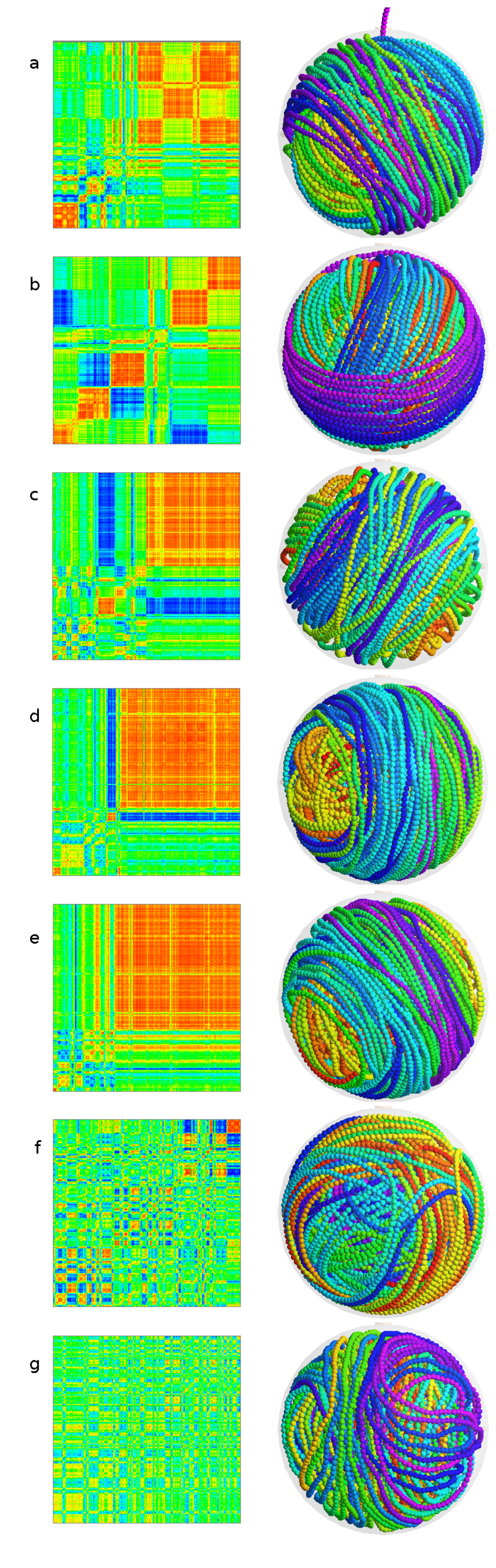}
\end{center}
\caption{\label{fig:4}Orientational correlation plots and pictures of selected chains;
details and color schemes are explained in the text (in the grayscale print version, dark
gray = blue/violet, medium gray = red, light gray = yellow/green); cases (a--e) are for
$f_a = 5000$, (f,g) are for $f_a = 3000$ and $1000$.}
\end{figure}

A more detailed explanation of how packaged chains are organized appears in
Fig.~\ref{fig:4}, where pictures of chain configurations are paired with orientational
correlation plots (explained below). The shell envelope is shown semi-transparently with
the portal at the top. Spectral colors are used for the chain configurations, ranging
from red at the chain head (the segments that enter first) to violet at the tail; thus
the image of a chain whose tail is adjacent to the shell wall typically has red segments
inside and violet loops on the outside, with the other colors interspersed. Note that
static pictures cannot reveal slow time-dependent behavior such as rotation of the
partially packaged chain, or spool rearrangement during and after insertion that can
obscure the final stages of the insertion history; animated image sequences provide
additional information.

Long chain sections arranged into spool-like configurations are a prominent feature of
the imagery, a mode of organization that can be quantified using orientational
correlation functions. These are evaluated by considering successive chain segments $i$
of length (e.g.) 40 ($1 \le i \le 200$) and, on the assumption that each segment is
approximately planar, evaluating its normal $\vec{n}_i$ as the mean of the cross
products of vectors between spheres in the segment spaced (e.g.) four apart. A
(symmetric) matrix $\textbf{C}$ is constructed, where $C_{ij} = \arccos (\vec{n}_i \cdot
\vec{n}_j)$ is the angle between the normals of segments $i$ and $j$, and displayed as a
2D plot with color denoting angular ranges (these colors unrelated to the pictures); the
chain head is in the lower-left corner. The $i$th row/column shows the alignment of
chain segments relative to $i$, where three segments amount to a single loop around the
shell. Regions of the chain with near-parallel alignment, the usual case for contiguous
segments that tend to be strongly correlated (except when an abrupt change occurs), are
shown in red, regions oriented antiparallel (to the original) in blue, and intermediate
cases in yellow and green.

The color plots do not show where individual segments are located in the shell, but by
comparing each $\textbf{C}$ with its picture it is apparent that a large upper-right red
square corresponds to a section of the chain forming a spool with multiple loops in
contact with, or close to, the shell wall; this, in turn, may be wrapped around one or
more differently aligned interior spools, typically terminating in a region with minimal
correlation extending to the chain head. Brief interruptions in the large red squares
correspond to short misaligned sections, sometimes just a single loop. The variety of
organizational patterns is unsurprising since the packed state represents the outcome of
a sequence of individual events that have little or no temporal relation; a typical
event might be a choice between a chain segment forcing previously packed contents to
rotate so that it can be accommodated on the exterior with minimal bending, or the
opening of a gap between loops so that it can penetrate to the interior.

Examples (a--e) are for the stiffest chains considered ($f_a = 5000$, the results for
$f_a = 4000$ are similar). Case (a) shows a chain just prior to complete insertion with
the final tail segment traversing the portal. Cases (b--e) show fully inserted chains
with increasingly large ordered regions. Each chain is packed differently; 22 of the 27
chains exceed 99\% completion (for this $f_a$), and in 12 of them the upper-right square
of $\textbf{C}$ is a red or red/yellow block that includes at least half the chain,
apart from a few defects, indicating that the dominant feature is a single spool. In
case (b) there are several smaller blocks corresponding to shorter spools,
concentrically arranged with differing orientations. In general, sets of spooled loops
with (near-) maximal radius form in (or close to) a median plane passing through the
portal; the spool axis direction varies slightly but can persist for a relatively long
time before undergoing sudden change. Previously inserted chain segments form a soft,
fluctuating `core' that rotates slowly and unevenly to accommodate new segments, usually
on or close to the outside. Note that twisting of the chain is not required, justifying
the omission of any torsional interaction; the minimal influence of torsion is
described in \cite{rol08}. The measured increasing mean radial distance
as a function of sphere location in the chain (Fig.~\ref{fig:3}) is consistent with
these observations.

The initial section of the chain to enter (up to $\approx 4000$ spheres, typically
exceeding the longest chains considered in previous work) also forms loops, although the
ordering is much weaker, and the loops are anisotropically compressed by subsequent
chain segments. This could indicate the existence of an interior `framework' that
assists spooling, implying a minimal chain length for spool development. The behavior is
the opposite of `inverse spooling' -- packing from the outside in -- seen in simulations
of shorter chains \cite{spa05,pet07,mah13}, even though concentric spools form in both
cases. The presence of multiple sets of spooled loops with
various orientations allows more uniform coverage of the shell surface and reduces the
bending energy (for both normal and inverse spools). The number of loops, assuming they
span the full circumference, is $l_b N / 2 \pi R_s \approx 70$.

The final two examples (f,g) are for more flexible chains, with $f_a = 3000$ and $1000$.
Packaging is less ordered at lower $f_a$, and loops with higher curvature appear more
often, so that while there is still some alignment, sizable spools are less likely. The
spheres also have reduced energetic preference as to initial placement after insertion,
so that the likelihood of the tail end of the chain being near the wall is lower
(Fig.~\ref{fig:3}).


In conclusion, the fact that the complex mechanisms employed by viruses for packaging
their genetic payloads are not readily understood has led to the development of reduced
models aimed at capturing the principal features. Simulations of such a model reveal
that spool-like organization appears spontaneously when packaging stiff polymer chains,
with the late-entering part of the chain preferably located close to the shell wall. The
efficacy of this packing scheme is obvious, in particular because the core rotation
alleviates the need to overcome chain twisting, although it has not been considered
previously. Apart from possible biological relevance, the results provide another
example \cite{rap14} of emergent cooperativity in the simplest of physical systems.

This work was initiated at the Aspen Center for Physics Workshop on Viral Assembly; the
hospitality of the Center and partial support from NSF grant PHYS-1066293 is
acknowledged.

\bibliography{polypack}

\end{document}